\documentclass[twoside,twocolumn,american,preprintnumbers,nofootinbib,aps,pra,10pt]{revtex4-1}

\pdfoutput=1
\usepackage{graphicx}
\usepackage{array}
\usepackage{tabularx}
\usepackage{float}
\usepackage{xcolor}
\usepackage{physics}
\usepackage{slashed}
\usepackage[font=small, labelfont=bf]{caption}
\usepackage[labelformat=simple]{subcaption}

\usepackage{amsmath, amsthm, amssymb, amsfonts}
\usepackage{multirow}
\usepackage[colorlinks= true, linkcolor =myblue , urlcolor = myblue, citecolor = myyellow,
anchorcolor = blue]{hyperref}

\definecolor{myblue}{rgb}{0.3, 0.5, 0.7}
\definecolor{myyellow}{rgb}{0.85, 0.72, 0.37}

\def\beq{\begin{align}}
\def\eeq{\end{align}}
\newcommand{\bi}{\begin{itemize}}
\newcommand{\ei}{\end{itemize}}
\newcommand{\ben}{\begin{enumerate}}
\newcommand{\een}{\end{enumerate}}
\newcommand{\be}{\begin{equation}}
\newcommand{\ee}{\end{equation}}
\newcommand{\bea}{\begin{eqnarray}}
\newcommand{\eea}{\end{eqnarray}}

\newcommand{\Kahler}{K\"ahler~}
\newcommand{\V}{\mathcal{V}}
\newcommand{\K}{\mathcal{K}}

\renewcommand{\O}{\mathcal{O}}

\newcommand{\de}{\partial}

\begin{document}

\title{Growth of Cosmic Strings beyond Kination}

\author{Luca Brunelli$^{1,2}$, Michele Cicoli$^{1,2}$, Francisco G. Pedro$^{1,2}$}
\affiliation{$^{1}$Dipartimento di Fisica e Astronomia, Universit\`a di Bologna, via Irnerio 46, 40126 Bologna, Italy}
\affiliation{$^{2}$INFN, Sezione di Bologna, viale Berti Pichat 6/2, 40127 Bologna, Italy}

\begin{abstract}
A novel mechanism for the production of a cosmic network of fundamental superstrings based on a time-varying string tension has been recently proposed in the context of a kinating background driven by the volume modulus of string compactifications. In this paper, we generalise the analysis of this growth mechanism by using dynamical system techniques. We first study the cosmological growth of strings in a spatially-flat Universe filled with a perfect fluid and a field-dependent tension, finding the fixed points of the phase space of this system. We then apply this analysis to  fundamental strings and EFT strings obtained from wrapping $p$-branes on $(p-1)$-cycles. We find a cosmological growth for fundamental strings even without kination, as in scaling fixed points, and for EFT strings arising from D3- and NS5-branes wrapped around fibration cycles.
\end{abstract}

\maketitle

\section{Introduction}
\label{Intro}

Cosmic strings are topological defects which arise generically in early universe cosmology from the spontaneous breaking of gauge symmetries \cite{Kibble:1976sj}. Interestingly, cosmic strings can form networks which lead to signals of new physics by emitting gravitational waves (see \cite{Vilenkin:2000jqa, Copeland:2009ga} for detailed reviews). 

A key property of cosmic strings is their tension which is usually considered as constant. However, from the point of view of a UV-complete theory like string theory, every feature of the low-energy effective theory is field-dependent, including the tension of cosmic strings. This observation opens up the possibility to have cosmic strings with a time-varying tension, a case which has already received a lot of attention \cite{Yamaguchi:2005gp,Ichikawa:2006rw,Cheng:2008ma,Sadeghi:2009wx,Wang:2012naa,Emond:2021vts}.

In particular, ref. \cite{Conlon:2024uob} exploited the time-dependence of the tension of fundamental strings to propose a novel mechanism for the formation of a cosmic network of these objects. In fact, the tension of fundamental strings depends on the value of the modulus controlling the size of the extra dimensions. In a background characterised by a kinating volume mode, as after the end of volume modulus inflation \cite{Conlon:2008cj,Cicoli:2015wja}, this tension can decrease in time causing the physical size of string loops to grow faster than the scale factor. If this epoch lasts long enough, an initial population of isolated small string loops will eventually percolate and form a network. The cosmological evolution of a string network with a time-varying tension has then been studied in \cite{Revello:2024gwa}. 

Note that fundamental strings are expected to be produced also at the end of brane-antibrane inflation \cite{Kofman:2005yz}. If they are stable, they can then form a cosmic network \cite{Sarangi:2002yt,Copeland:2003bj}. As recently shown in \cite{Cicoli:2024bwq}, a proper stabilisation of the volume modulus during brane-antibrane inflation can lead to a scenario where, after the end of inflation, the volume modulus rolls towards its late-time minimum from much smaller field values. While rolling, the modulus can lead to particle production which could also be sourced by the cosmic string network. This justifies the picture of an expanding Universe with a rolling modulus, a fluid and a cosmic string network with a time-dependent tension.

The goal of this paper is to extend the study of the growth mechanism proposed in \cite{Conlon:2024uob} by exploring two possible generalisations: ($i$) a dynamics beyond kination, for a spatially-flat Universe filled not just with a volume modulus but also with a perfect fluid; ($ii$) strings beyond fundamental ones, as effective strings obtained by wrapping $p$-branes on $(p-1)$-cycles.

The analysis of the corresponding dynamical system and its fixed points reveals a few interesting facts: 
\begin{enumerate}
\item Kination is an unstable fixed point. This implies that the physically relevant situation, i.e. the one of a fast-rolling modulus corresponding to a small deviation from the kination fixed point, can at most be a transient. However, we estimate that under general conditions one can still allow for a transient of $\mathcal{O}(10)$ e-foldings where string loops grow in size.

\item The growth of the physical size of cosmic strings can occur even without kination, as in scaling fixed points where the energy density of the modulus tracks that of background fluid. Contrary to kination, the scaling fixed point is stable, allowing for a very long epoch where the growth condition is satisfied. 

\item The scaling fixed point exists and yields a growth for particular values of the power of the volume in the scalar potential and the equation of state parameter of the fluid.

\item String compactifications are in general characterised by several moduli beyond the volume mode. If the rolling field is a bulk modulus different from the volume, the growth condition can be satisfied for several cases of effective strings. In the type IIB case, these objects would originate from D3-branes wrapped around 2-cycles or NS5-branes wrapped around 4-cycles.

\item When EFT strings arise from branes wrapped around blow-up modes, a string growth could in principle occur but its duration is extremely short since blow-up modes are characterised by sub-Planckian field ranges which cannot sustain a long-lasting epoch of kination.
\end{enumerate}

\section{Strings with time-dependent tension}
\label{SecII}

In this section we briefly review the time evolution of a string with time-dependent tension following \cite{Emond:2021vts,Conlon:2024uob,Revello:2024gwa}.
The Nambu-Goto (NG) action for a string with spacetime-dependent tension $\mu(x)$ is given by:
\begin{equation}
\label{eq:NG string}
S_{\rm NG}=-\int \dd^2 \sigma \, \mu(x) \,  \sqrt{-\gamma}\,, 
\end{equation}
where $x^\mu$ are the spacetime coordinates, $\sigma^a$ the worldsheet coordinates and $\gamma$ the determinant of the induced metric $\gamma_{ab}$ defined as: 
\begin{equation}
\gamma_{ab}= \frac{\partial x^\mu}{\partial \sigma^a} \frac{\partial x^\nu}{\partial \sigma^b}\, g_{\mu\nu}\,.
\end{equation}
Varying the NG action yields the string equations of motion \cite{Emond:2021vts}:
\begin{equation}
    \frac{1}{\mu \sqrt{-\gamma}}\, \partial_a(\mu \sqrt{-\gamma}\gamma^{ab}x^\mu_{,b})+\Gamma^\mu_{\nu\rho} \gamma^{ab}x^\nu_{,a}x^\rho_{,b}-\frac{\partial^\mu \mu}{\mu}=0\,.
\end{equation}
Choosing coordinates such that $x^0=\sigma^0$ and the condition $\frac{\partial \vec{x}}{\partial \sigma^1}\cdot\frac{\partial\vec{x}}{\partial \sigma^0}=0$ holds, and assuming a closed circular string in the $z=0$ plane: 
\begin{equation}
(x(t),y(t),z(t))= R(t)(\cos \theta, \sin \theta,0)\,,
\end{equation}
in a flat FLRW spacetime, the equations of motion become:
\begin{align}
    \frac{\dot{\epsilon}}{\epsilon} &=H-a^2 \dot{R}^2\left(2 H+\frac{\dot{\mu}}{\mu}\right),
\label{eq:epsilonEOM}\\
\ddot{R}+ H \dot{R}\,+& \frac{R}{\epsilon^2}+\left(2 H+\frac{\dot{\mu}}{\mu}\right )(1-a^2 \dot{R}^2)\dot{R}=0\,,
\label{eq:REOM}
\end{align}
where we defined $\epsilon\equiv a |R| / \sqrt{1-a^2 \dot{R}^2}$.

Considering the constant tension and flat spacetime limit of \eqref{eq:epsilonEOM}-\eqref{eq:REOM} ($H=0$, $\mu$ and $a$ constant), one automatically finds the solution $R=R_{\rm max} \cos (t/\epsilon )$ with $\epsilon$ constant. This prompts the identification of $\epsilon$ with the oscillation period. Simultaneously, the definition of $\epsilon$ allows for its identification with the maximum physical radius of the oscillating string: $\epsilon (t)= a(t) R_{\rm max}(t)$.

\begin{figure}
    \centering
    \includegraphics[width=0.8\linewidth]{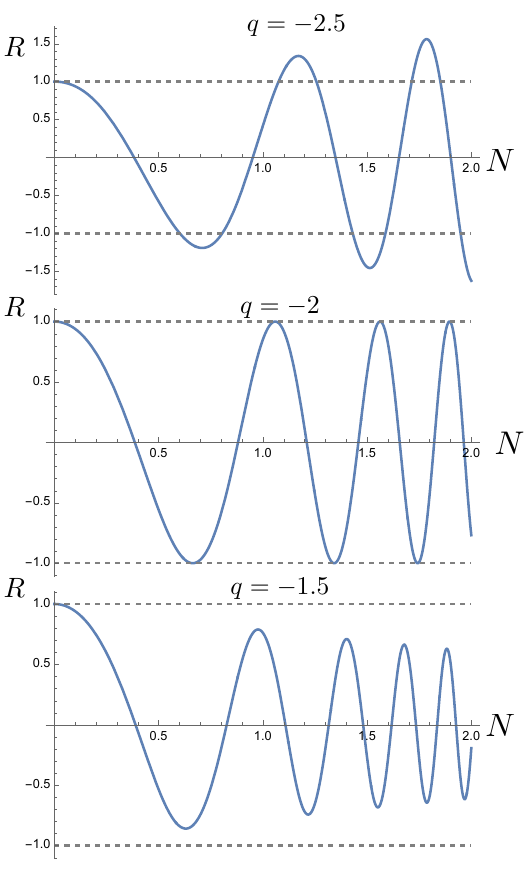}
    \caption{Time evolution of the string comoving radius in matter domination for $\mu\propto a^q$ with $q=\{-2.5, -2, -1.5\}$.}
    \label{fig:RN}
\end{figure}

For a string in an expanding spacetime there are two relevant time-scales: its period $T_{\rm str}=\epsilon$, and the Hubble time $t_H=1/H$. If $\epsilon\ll 1/H$, then averaging over many oscillations in a Hubble time yields $\langle a^2 \dot{R}^2\rangle=1/2$. In this regime the string is small when compared with the horizon size, $R_{\rm max}\ll H$, and we may approximate:
\begin{equation}
\frac{\dot{\epsilon}}{\epsilon}\simeq -\frac{1}{2}\frac{\dot{\mu}}{\mu}\,.
\end{equation}
This implies that $\epsilon$, and consequently $R_{\rm max}$, grows if the string tension decreases with time. In general the string grows if and only if the RHS of \eqref{eq:epsilonEOM} is positive, which is guaranteed to happen if:
\begin{equation}
\label{eq:growth condition}
    2H+\frac{\dot{\mu}}{\mu}<0\,,
\end{equation}
or, in other words, if the tension falls faster than $a^{-2}$, regardless of the fluid dominating the energy momentum tensor. In Fig. \ref{fig:RN} we show the evolution of the string comoving radius over two e-foldings of expansion in a matter dominated background.

\section{Cosmological evolution and string growth}
\label{sec:exp. tension}

Throughout this work the time dependence of the string tension is given by the dynamics of a scalar field $\phi$. Let us consider the case in which $\mu$ depends exponentially on such a scalar:
\begin{equation}
\label{eq:exp. mu}
    \mu= M^2 \, e^{-\xi \,\phi/M_p}\,,
\end{equation}
where $\xi$ is a dimensionless constant and $M$ a parameter with the dimension of a mass. The growth condition \eqref{eq:growth condition} can be rephrased by switching from cosmological time $t$ to the number of e-foldings $N = \ln a$ as (setting $M_p=1$):
\begin{equation}
\label{eq:gc xi-phi'}
2-\xi\, \phi'<0\,,
\end{equation}
where $'\equiv d/d N$. Following \cite{Copeland:1997et}, we define the field phase space variables:
\begin{equation}
\label{eq:x&y efolds}
X \equiv \frac{\dot \phi}{\sqrt{6}\, H} = \frac{\phi'}{\sqrt{6}}\, , \qquad Y \equiv \frac{1}{H}\sqrt{\frac{V(\phi)}{3}}\,,
\end{equation}
where we focus on $V(\phi)\geq 0$. In this way the growth condition for an exponential string tension can be recast as a constraint on the phase space:
\begin{equation}
\label{eq:gc x}
    2- \sqrt{6}\,  \xi \, X <0\, .
\end{equation}
In what follows we will determine when this condition is satisfied, assuming that the energy density of the cosmic strings is negligible compared to $H^2$.  

We will consider the evolution of the field $\phi$ in a spatially-flat Universe filled with a perfect barotropic fluid with an equation of state:
\begin{equation}
\label{eq:eq of state}
    p_{\rm f} = \omega\,  \rho_{\rm f}\,,
\end{equation}
with $-1 \leq \omega \leq 1$. The Friedman constraint:
\begin{equation}
\label{eq:friedmann constraint}
H^2 = \frac{1}{3} \left(\rho_{\rm f} +\frac12 \dot \phi^2 + V(\phi)\right),
\end{equation}
can be written in terms of $X$ and $Y$ as:
\begin{equation}
\frac{\rho_{\rm f}}{3 H^2} + X^2+ Y^2 = 1\,.
\end{equation}
This implies that the field phase space is constrained within the region:
\begin{equation}
\label{eq:phase space}
    X^2 + Y^2 \leq 1\,.
\end{equation}
Given that we restrict our analysis to the case of an expanding Universe with $H>0$, (\ref{eq:x&y efolds}) implies that the phase space available for the field $\phi$ is the upper-half unitary disk with $Y\geq 0$. 

\begin{figure}[h]
\centering
\raisebox{-0.5\height}{\includegraphics[width=0.5\textwidth]{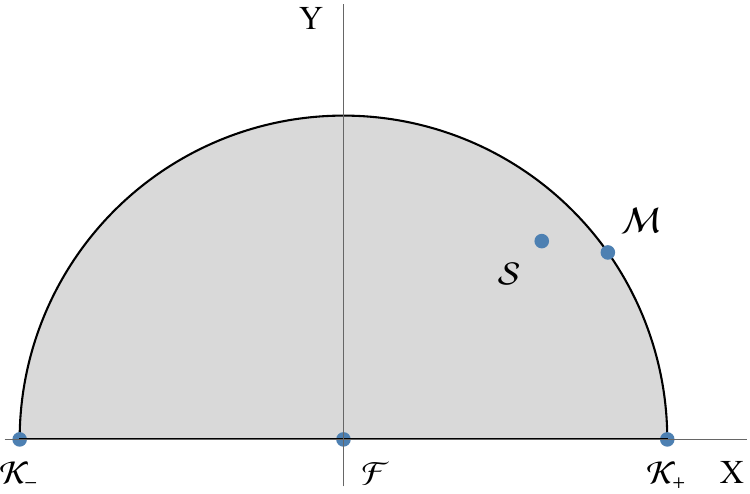}}
\caption[Short Caption]{\hspace{0pt} \captionsize Phase space of the autonomous system \eqref{eq:X'}-\eqref{eq:Y'} describing the evolution of $\phi$. The blue dots denote the fixed points for the illustrative example with $\lambda=2$ and $\omega=0$.}
\label{fig:ps}
\end{figure}

In order to derive exact analytic results, we will follow \cite{Copeland:1997et} and assume an exponential potential for $\phi$:
\begin{equation}
\label{eq:exp potential}
    V(\phi) = V_0\, e^{-\lambda \phi}\, ,
\end{equation}
with $\lambda$ constant. This is a rather natural choice for some string moduli (see Sec. \ref{sec:volume}, \ref{sec:dilaton}, \ref{sec:fibre}) while it is not for others (Sec. \ref{sec:blowup}). We will highlight the key differences in that case, and adapt the methods we develop in this section. For a potential of the form \eqref{eq:exp potential}, the Friedman and Klein-Gordon equations form an autonomous system which can be written in terms of $X$ and $Y$ as \cite{Copeland:1997et}:
\begin{align}
X' & = -3X + \lambda \sqrt{\frac{3}{2}} Y^2+ \frac{3}{2}X\left[ 2X^2 + (\omega +1) (1-X^2-Y^2)\right] \label{eq:X'}\\
Y' & = - \lambda \sqrt{\frac{3}{2}}\, XY + \frac{3}{2} Y \left[ 2X^2 + (\omega +1) (1-X^2-Y^2)\right]. \label{eq:Y'}
\end{align}
The fixed points of the system \eqref{eq:X'}-\eqref{eq:Y'} are known and can be classified. We show the phase space for $X$ and $Y$ with all possible fixed points in Fig. \ref{fig:ps}.

$\K_\pm$, located at the points $(\pm 1, 0)$ in the $(X,Y)$-plane, are the \textit{kination} fixed points where all the energy density of the Universe is given by the kinetic energy of the field $\phi$. $\mathcal{F} = (0,0)$ is the \textit{fluid domination} point where the energy density of the Universe is thoroughly dominated by $\rho_{\rm f}$. $\mathcal{M}$ and $\mathcal{S}$ denote two families of fixed points parametrised by $\lambda$ and $\omega$. $\mathcal{M}$ is the \textit{modulus domination} family of fixed points, in which $\phi$ dominates the energy density of the Universe, so that $X^2+ Y^2 = 1$. In particular, in the $\lambda \to 0$ limit, this fixed point would correspond to pure dS with $\mathcal{M}=(0,1)$. On the other hand, $\mathcal{S}$ denotes \textit{scaling} fixed points. In this regime the energy density of the field redshifts as the one of the fluid. 

We now want to analyse the growth condition \eqref{eq:gc x} and retrieve some bounds on the parameters at play simply by combining it with the existence condition of the fixed points of our autonomous system \eqref{eq:X'}-\eqref{eq:Y'}. First of all, rearranging \eqref{eq:gc x} we get:
\begin{equation}
\label{eq:explicit gc}
|X| > \sqrt{\frac{2}{3}} \frac{1}{|\xi|}\, .
\end{equation}
Imposing $|X|\leq 1$ on \eqref{eq:explicit gc} we get a lower bound on $|\xi|$:
\begin{equation}
\label{eq:lower bound on xi}
|\xi| > \sqrt{\frac{2}{3}} \simeq 0.816\,.
\end{equation}
As we will see in the following sections, $\xi$ is fixed by the canonical normalisation of the scalar field controlling the time-dependence of the string tension $\mu$. Let us now see if we can push some of the fixed points within the growth region for the physical size of the string given by \eqref{eq:explicit gc}. In Tab. \ref{tab:gc fixed points} we collect the main results obtained by combining \eqref{eq:explicit gc} with the existence conditions for each fixed point.

\begin{table*}[t!] 
\centering
    \begin{tabular}{| m{1cm} | m{2.1cm}| m{2.5cm} |m{3cm}| m{8cm}|}
        \hline
        \centering \textbf{FP} &\centering \textbf{X} & \centering \textbf{Y} & \centering \textbf{Existence} &  $\qquad$ \textbf{Existence and Growth when} $|\xi|>\sqrt{2/3}$ \\
        \hline
        \centering $\K_+$ & \centering 1 & \centering 0 & \centering $\forall  \lambda\,$ and $\,\forall \omega $ & $\qquad \qquad \qquad \qquad\quad \forall  \lambda\,$ and $\,\forall \omega$  \\
        \hline
        \centering $\K_-$ & \centering  -1 & \centering 0 & \centering  $\forall  \lambda\,$ and $\,\forall \omega$ & $\qquad \qquad \qquad \qquad\quad\forall \lambda\,$ and $\, \forall \omega$  \\
       \hline
      \centering $\mathcal{F}$  & \centering 0 &  \centering 0 & \centering $\forall  \lambda\,$ and $\, \forall \omega $   & $\qquad \qquad \qquad \qquad \qquad \,  $never \\
      \hline
       \centering $\mathcal{M}$ &\centering  $\frac{\lambda}{\sqrt{6}}$ &\centering  $\sqrt{1-\frac{\lambda^2}{6}}$ & \centering $|\lambda|< \sqrt{6}$   & $\qquad \qquad \qquad \qquad $ $\frac{2}{|\xi|}<|\lambda|< \sqrt{6}$  \\
        \hline 
        \centering $
        \mathcal{S}$ & \centering $\sqrt{\frac{3}{2}} \frac{\omega+1}{\lambda}$ & \centering  $\sqrt{\frac{3(1-\omega^2)}{2 \lambda^2}}$ & \centering $\lambda^2 >3(\omega +1)$ & $\quad\sqrt{3(\omega+1)}<|\lambda|<\frac{3}{2} (\omega+1) |\xi|\,\,$  with $\,\,\sqrt{\omega+1} > \frac{2}{\sqrt{3} |\xi|}$ \\
        \hline
    \end{tabular}
    \caption{Fixed points (FP) of the autonomous system \eqref{eq:X'}-\eqref{eq:Y'} with existence and growth conditions.}
    \label{tab:gc fixed points}
\end{table*}

Clearly, if the growth region exists at all within the phase space, as per \eqref{eq:lower bound on xi}, one of the kination fixed points will be contained in it. This is intuitively true because the growth condition depends on the `velocity' of the field  $\phi$, and kination is the fixed point of maximal field velocity. Hence, if kination is not enough for the growth condition to be satisfied, no other point in the phase space can satisfy \eqref{eq:gc x}. Equally intuitive is the fact that $\mathcal{F}$ is never going to be within the growth region. Indeed, in that case, the dynamics of the field $\phi$ plays no role at all in the evolution of the Universe. 

The cases of the fixed points $\mathcal{M}$ and $\mathcal{S}$ are more subtle. An interesting aspect of these is that, unlike kination, they can be \textit{stable}. In particular, if $\mathcal S$ exists within the growth region,
it is stable \cite{Copeland:1997et}. This meansthat the system can stay in a scaling regime for as long as \eqref{eq:exp potential} is a good approximation for the scalar potential. The large number of e-foldings that the system spends in $\mathcal S$ ensures a long period of growth of cosmic strings. The $\mathcal S$ existence condition and string growth condition combine to impose constraints on combinations of $\lambda$, $\omega$ and $\xi$. The growth condition \eqref{eq:explicit gc} can be rephrased in an upper bound for $\lambda$ depending on $\omega$ and $\xi$, while the existence condition furnishes a lower bound:
\begin{equation}
\label{eq:range for lambda scaling}
    \sqrt{3(\omega+1)}<|\lambda| < \frac{3}{2} (\omega + 1) |\xi|\, .
\end{equation}
This interval is guaranteed to exist if:
\begin{equation}
\label{eq:existence condition scaling growth}
    \sqrt{\omega + 1} > \frac{2}{\sqrt{3}\, |\xi|}\, .
\end{equation}
In the following sections we will fix $\xi$ from UV physics and use \eqref{eq:existence condition scaling growth} to constrain what fluids can give a scaling regime within the growth region of the phase space. 

The modulus dominated fixed point $\mathcal M$, when it exists, can lie within the growth region provided that the potential is sufficiently steep. In fact, the existence condition for $\mathcal M$ and \eqref{eq:explicit gc} combine to give:
\begin{equation}
\label{eq:gc field dom}
    \frac{2}{|\xi|} < |\lambda| < \sqrt{6}\, .
\end{equation}
Such a parameter region always exists if \eqref{eq:lower bound on xi} is satisfied.

\section{Fundamental strings}
\label{sec:fundamental Strings}

We now turn to analysing the dynamics of the field $\phi$ from a string theory point of view, focussing first on the case of fundamental strings. As mentioned already in Sec. \ref{Intro}, a volume modulus fast-rolling towards its late-time minimum is a common post-inflationary feature of several string inflation models, like volume modulus inflation \cite{Conlon:2008cj,Cicoli:2015wja} and brane-antibrane inflation \cite{Cicoli:2024bwq}. Particle production from the evolving volume mode or from a cosmic string network formed after brane-antibrane annihilation justifies the presence of a barotropic fluid in the background. The study of the condition for the growth of the physical size of the fundamental strings is then important to check whether an initial population of isolated small loops can percolate and form a network, or how the cosmological evolution and gravity wave emission of an already formed network (see \cite{Avgoustidis:2025svu}) can be affected by the growth. 

Let us start by noticing that the length of fundamental strings is proportional to the string length $\ell_s =2\pi \sqrt{\alpha'}$, and so their tension $\mu$ is proportional to the square of the string scale $\mu \sim M_s^2$. In Einstein frame, we can express $M_s$ in terms of the 4D Planck mass $M_p$ as:
\begin{equation}
\label{eq:string scale}
    M_s = \frac{g_s^{1/4} M_p}{\sqrt{4 \pi \V}}\,.
\end{equation}
The value of $M_s$ is controlled by two scalar fields: the volume modulus $\V$ and the dilaton $\varphi$ which determines the strength of the string coupling $g_s$. In the following sections we study the evolution of the string tension $\mu$ depending on the dynamics of these fields.

\subsection{Volume mode}
\label{sec:volume}

If we suppose that only the volume modulus evolves in time while the dilaton is fixed, the relevant moduli dependence of the string tension is given by:
\begin{equation}
\label{eq:tension wrt volume}
    \mu \sim \frac{M_p^2}{\V}\, .
\end{equation}
Working within type IIB compactifications, one can find the canonical normalisation of the volume modulus from the \Kahler potential:
\begin{equation}
    \frac{K}{M_p^2} = -2 \ln \V\,.
\end{equation}
In the simplest case, one can express the volume simply as:
\begin{equation}
\label{eq:vol(tau)}
\V  = \tau^{3/2}\,,
\end{equation}
where $\tau$ is the volume of a 4-cycle. Hence, the kinetic Lagrangian for the field $\tau$ reads:
\begin{equation}
\mathcal{L}_{\rm kin} \supset \frac{3 M_p^2}{4\tau^2}\,\de_\mu \tau \de^\mu \tau\,,
\end{equation}
and the canonical normalisation of $\tau$ turns out to be:
\begin{equation}
\label{eq:can. norm. V}
\frac{\chi}{M_p} = \sqrt{\frac{3}{2}}\ln \tau =\sqrt{\frac{2}{3}} \ln \V\,.
\end{equation}
Therefore, from \eqref{eq:tension wrt volume} the dependence of $\mu$ on $\chi$ is exponential:
\begin{equation}
\label{eq:tension wrt Phi}
\mu \sim M_p^2\, e^{-\sqrt{\frac{3}{2}} \chi/M_p}\,.
\end{equation}
Comparing this expression with (\ref{eq:exp. mu}), we set:
\begin{equation}
\label{eq:xi volume}
\xi = \sqrt{\frac{3}{2}}\,.
\end{equation}
To use the methods outlined in Sec. \ref{sec:exp. tension}, the scalar potential should take the exponential form (\ref{eq:exp potential}). This is indeed the case for the volume mode since, in the large volume limit where the EFT is under control, its potential is dominated by power-law contributions arising from perturbative corrections to the K\"ahler potential: 
\begin{equation}
\label{eq:typical V pot.}
V \simeq \frac{V_0}{\V^p}\,,
\end{equation}
with $V_0$ depending on details of the compactification. From \eqref{eq:can. norm. V}, we see this is indeed of the form \eqref{eq:exp potential}, with $\lambda = \xi \,p$. 

The first thing to note is that $\xi=\sqrt{3/2}$ satisfies the existence condition (\ref{eq:lower bound on xi}) for the growth mechanism and the growth region in the phase space \eqref{eq:explicit gc} becomes:
\begin{equation}
\label{eq:volume gc}
X>\frac{2}{3}\,.
\end{equation} 
Let us now analyse each fixed point separately.

\paragraph*{\textbf{Kination:}}

The kination fixed point $\K_+$ lies within the growth region. The physical case of a fast-rolling volume mode would not correspond exactly to $\K_+$ since in $\K_+$ the potential vanishes. However it would correspond to initial conditions very close to $\K_+$. As stated above, $\K_+$ is an unstable fixed point, and so the late-time attractor of the system will not be $\K_+$. However, if the system starts to evolve with an initial condition very close to $\K_+$, it will spend a long time in the growth region \eqref{eq:volume gc}, even if the fixed point is unstable. In fact, it is not strictly necessary that the late-time attractor solution lies within the growth region, as long as the system spends a sufficiently long amount of time within the growth region in a transient state. 

\begin{figure}
\centering
\includegraphics[width=0.5\textwidth]{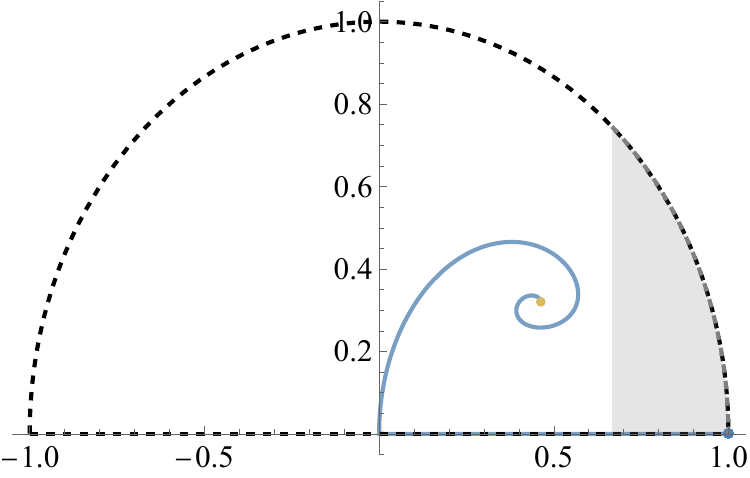}
\caption{Phase space evolution of the system with initial conditions $X_i = 1- 10^{-6}$ and $Y_i = 10^{-3}$ (blue dot), towards the scaling fixed point (yellow dot) for $\omega = 1/3$ and a potential of the form \eqref{eq:typical V pot.} with $p = 3$. The shaded area represents the growth region.}
\label{fig:evolution}
\end{figure}

In Fig. \ref{fig:evolution}, we show the evolution of the autonomous system for the initial condition $X_i = 1-10^{-6}$ and $Y_i=10^{-3}$, a radiation fluid ($\omega=1/3$) and a potential of the form (\ref{eq:typical V pot.}) with $p = 3$, as typical for a volume modulus. Since $\K_+$ is an unstable fixed point, it does not act as an attractor, and the system eventually moves away from it leaving the growth region in phase space represented by the shaded area and reaching the scaling fixed point.

\begin{figure}
\centering
\includegraphics[width=0.5\textwidth]{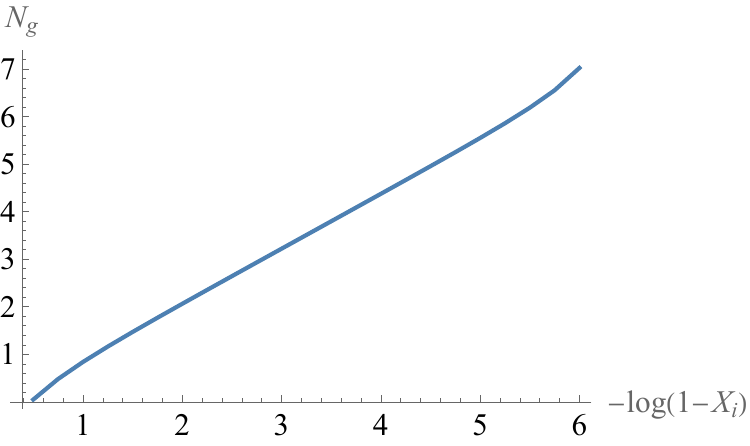}
\caption{Number of e-foldings of growth $N_g$ as a function of the initial condition $X_i$ (given in terms the logarithm of the distance from $\K_+$) for $\omega = 1/3$ and $p = 3$.}
    \label{fig:Ng-Xi}
\end{figure}

However, very close to $\K_+$, $X'$ is tiny, and so it takes a few e-foldings for the system to exit the growth region. A plot of the number of e-foldings of growth $N_g$ as a function of the initial condition $X_i$ is shown in Fig. \ref{fig:Ng-Xi}.

Whether this number of e-foldings is enough for the strings to percolate into a cosmic network depends on the initial length and number density of the population of fundamental strings. The complete assessment of this analysis is a very interesting direction for the future research. Note that our analysis is consistent with the findings of \cite{Conlon:2024uob,Revello:2024gwa}.

\paragraph*{\textbf{Modulus domination:}}
Using the results summarised in Tab. \ref{tab:gc fixed points} with the value of $\xi$ given by \eqref{eq:xi volume}, we see that, in addition to $\K_+$, for certain values of $\lambda$ also $\mathcal M$ can lie within the growth region. In particular, the point $\mathcal M$ exists and satisfies the growth condition whenever:
\begin{equation}
\label{eq:lambda C volume}
2 \sqrt{\frac{2}{3}}< \lambda < \sqrt{6}\,,
\end{equation}
corresponding to a potential \eqref{eq:typical V pot.} with a volume scaling:
\begin{equation}
\label{eq:p C volume}
\frac{4}{3} < p < 2 \,.
\end{equation}

\begin{figure}
    \centering
    \includegraphics[width=0.48\textwidth]{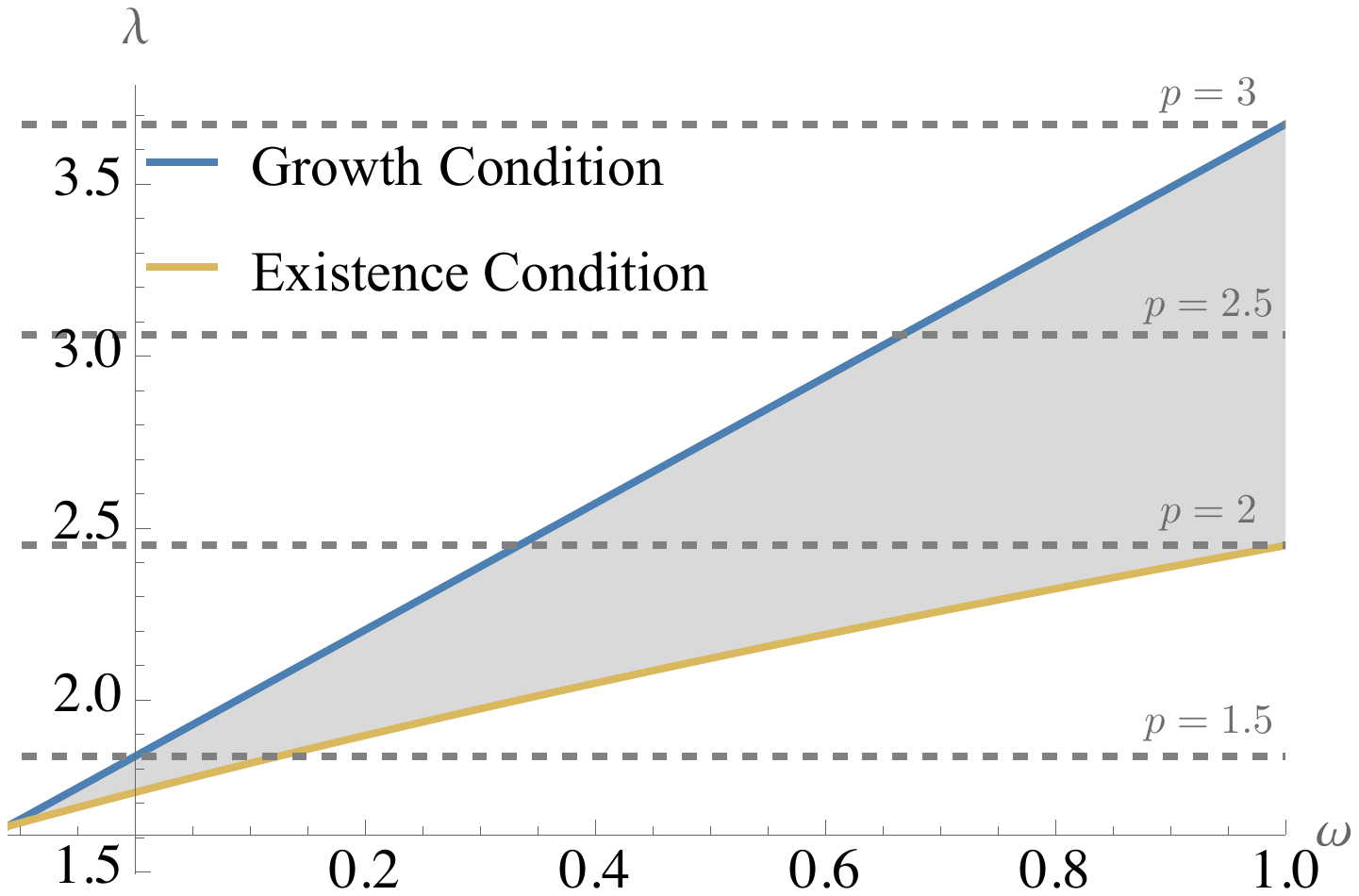}
    \caption{Range of $\lambda$'s for which the scaling fixed point lies within the growth region as a function of $\omega$. The dashed lines mark reference values for $p$.}
    \label{fig:lambda-omega}
\end{figure}

\paragraph*{\textbf{Scaling fixed point:}}
Contrary to $\mathcal{M}$, the existence and growth conditions for the point $\mathcal S$ do not depend just on $\xi$ and $\lambda$, but also on the equation of state parameter of the barotropic fluid. In full generality, $\mathcal{S}$ lies in the growth region whenever:
\begin{equation}
\label{eq:general lambda S}  
 \sqrt{3 (\omega +1)}< \lambda < (\omega+1) \left(\frac{3}{2}\right)^{3/2}\,,
\end{equation}
corresponding to a range:
\begin{equation}
\label{eq:range p S}
    \sqrt{2(\omega+1)}<p<\frac{3}{2}(\omega+1)\,.
\end{equation}
This window is open whenever:
\begin{equation}
\label{eq:omega S}
\omega >-\frac{1}{9}\,.
\end{equation}
We plot this range of $\lambda$ with respect to $\omega$ in Fig. \ref{fig:lambda-omega}. For example, $\mathcal S$ is in the growth regime in the presence of radiation ($\omega=1/3$) if $\sqrt{8/3}<p<2$, and in the presence of matter ($\omega=0$) if $\sqrt{2}<p<3/2$.

\begin{figure}
\centering
\includegraphics[width=0.5\textwidth]{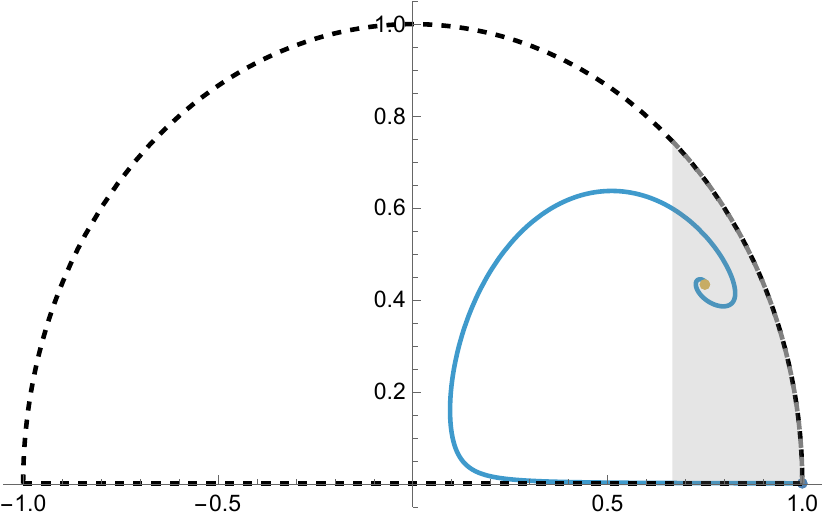}
\caption{Phase space evolution of the system with initial conditions $X_i = 1- 10^{-6}$ and $Y_i = 10^{-3}$ (blue dot), towards the scaling fixed point (yellow dot) for $\omega = 1/2$ and a potential of the form \eqref{eq:typical V pot.} with $p = 2$. The shaded area represents the growth region.}
\label{NewEvol}
\end{figure}

For illustrative purposes, we choose the case with $p=2$ and $\omega=1/2$ where $\mathcal{S}$ is fully inside the growth region and plot it in Fig. \ref{NewEvol} choosing the same initial conditions as in Fig. \ref{fig:evolution}. In this case there would be two epochs of string growth: the first during the initial transient close to $\K_+$ and the second, much longer, when the system reaches the final attractor $\mathcal{S}$. Being $\mathcal{S}$ an attractor, the system remains there until the volume mode gets close to the minimum of its potential where $V$ can no longer be approximated by a single contribution as in (\ref{eq:typical V pot.}).

\subsection{Dilaton}
\label{sec:dilaton}

We now consider the case in which the dilaton field controlling the string coupling is not fixed but it is rolling along its potential. From \eqref{eq:string scale}, we can retrieve the dependence of $\mu$ on $\varphi$:
\begin{equation}
\mu \sim M_s^2 \sim \frac{M_p^2}{\mathcal{V}} g_s^{1/2}\simeq \frac{M_p^2}{\mathcal{V}} e^{\frac{\varphi}{2}}\,,
\end{equation}
where we used $g_s = e^\varphi$ to express the string coupling in terms of the dilaton $\varphi$. Considering the volume as fixed,\footnote{One could also consider the case in which both fields are evolving. This is not impossible since in some flux compactifications the dilaton is light \cite{Chauhan:2025rdj}. However, in this case the time-dependence of the string tension becomes more complicated. We leave this analysis for future work.} the tension is again of the form \eqref{eq:exp. mu} with: 
\begin{equation}
\label{eq:xi dilaton}
    \xi = \frac{1}{2}\,.
\end{equation}
However, this factor does not satisfy the condition \eqref{eq:lower bound on xi}, meaning that there is no growth region within the phase space. We are therefore led to the conclusion that a rolling dilaton can never drive the growth of a fundamental string.

\section{EFT strings}
\label{sec:eft strings}

The fundamental strings described in Sec. \ref{sec:fundamental Strings} are not the only kind of 1D objects present in the 4D EFT of string theory. String-like objects in spacetime can originate also from $p$-branes wrapping $(p-1)$-cycles in the internal dimensions. We name these EFT strings since they are products of the 4D EFT. Considering this case is interesting because the string tension can decouple from the string scale. In fact, the tension $\mu$ can be obtained by reducing the world-volume action of a $p$-brane on a $(p-1)$-cycle $\Sigma_{p-1}$. Indeed, the action of a $p$-brane reads:
\begin{equation}
\label{eq:worldvolume action}
S_p \simeq -T_p \int_{\mathcal W_{p+1}} \sqrt{-\det(g_{\alpha \beta})} (1+\dots)\, \dd^{p+1} \zeta\,,
\end{equation}
where $T_p \sim M_s^{p+1}$ is the tension of the brane, the dots indicate extra terms which depend on the specific kind of brane, and $g_{\alpha\beta}$ is the pull-back of the metric on the world-volume $\mathcal{W}_{p+1}$ which can be factorised as:
\begin{equation}
\mathcal W_{p+1} = \mathcal W_2\,\times  \Sigma_{p-1}\,,
\end{equation}
in terms of the world-sheet of the non-compact dimensions $\mathcal W_2$ and the $(p-1)$-cycle $\Sigma_{p-1}$ wrapped by the $p$-brane. Hence $S_p$ factorises as:
\begin{equation}
S_p \supset -T_p \int_{\mathcal W_2} \dd^2 \sigma \, \sqrt{-\det(g_{ab})} \, \int_{\Sigma_{p-1}} \dd^{p-1} \zeta \,\sqrt{\det (g_{i j})}\,.
\label{Sp}
\end{equation}
The second term yields the volume of $\Sigma_{p-1}$ in string units $\text{Vol}(\Sigma_{p-1})$. Then, the action \eqref{Sp} takes the same form as \eqref{eq:NG string} with:
\begin{equation}
\label{eq:tension of EFT string}
\mu \sim \text{Vol}(\Sigma_{p-1})\, M_s^{p+1}\,.
\end{equation}
In type IIB compactifications on Calabi-Yau (CY) orientifolds with O3/O7-planes, it is natural to consider D3-branes wrapped around 2-cycles of volume $t$ in string units, and NS5-branes wrapped around 4-cycles of volume $\tau$. Indeed, duality arguments indicate that an NS5-brane should have a DBI-like action containing the tension of the brane \cite{Eyras:1998hn} which can be factorised as in \eqref{eq:worldvolume action}.

Now, as already noted in \cite{Conlon:2024uob}, if $t$ is the 2-cycle dual to the 4-cycle $\tau$ controlling the CY volume as in \eqref{eq:vol(tau)}, such that $t \propto \sqrt{\tau}\simeq \V^{1/3}$, the tension is given by:
\begin{equation}
\label{eq:mu eft volume}
\mu \sim \frac{M_p^2}{\V^{2/3}} = M_p^2 \, e^{-\sqrt{\frac{2}{3}} \chi/M_p}\,.
\end{equation}
This falls in the same exponential classification as before, but now with:
\begin{equation}
\label{eq:xi eft volume}
\xi = \sqrt{\frac{2}{3}}\,,
\end{equation}
which is marginally outside the growth regime given by \eqref{eq:lower bound on xi}. This means that, even for $X=1$, the condition \eqref{eq:explicit gc} is not satisfied, and so the string does not grow in size during the evolution of the Universe, behaving as if it were in flat spacetime \cite{Emond:2021vts}. The situation gets even worse if we wrap an NS5-brane around the 4-cycle $\tau$. In that case, from \eqref{eq:tension of EFT string}, $\mu \sim M_p^2\, \V^{-1/3}$ which yields $\xi < \sqrt{2/3}$.

Therefore, in the search for the growth of EFT strings, we are led to consider situations where the tension depends not just on $\mathcal{V}$ but also on cycles different from the overall volume. We will then consider the volume as fixed during the evolution of other cycles.\footnote{The opposite situation, where $\mathcal{V}$ evolves while the orthogonal moduli are fixed, does not lead to any string growth since $\xi$ would take the same values as for D3-branes wrapped on $t$ and NS5-branes wrapped on $\tau$ for the single modulus case.} It is the canonical normalisation of these K\"ahler moduli which determines $\xi$. To be explicit, we split this analysis into two subcases:
\begin{enumerate}
\item[(a)] Bulk cycles in situations when the CY volume is controlled by more than one cycle, as in K3 or $T^4$-fibred CY threefolds.
\item[(b)] Blow-up cycles resolving point-like singularities, like exceptional del Pezzo divisors.
\end{enumerate}
These two cases turn out to be substantially different given that the form of the canonical normalisation of bulk cycles is exponential, while that of blow-up modes is monomial. In particular, this means that in the former case we can directly apply the methods of Sec. \ref{sec:exp. tension}, while in the latter the growth condition takes a different form.

\subsection{Bulk cycles in fibred Calabi-Yaus}
\label{sec:fibre}

Let us now focus on the case of EFT strings which are branes wrapped on bulk cycles different from the volume mode. For illustrative purposes but without loss of generality, we shall focus on the simplest case where the CY threefold is a K3 or $T^4$ fibration over a $\mathbb{P}^1$ base with volume $\V$ in string units given by \cite{Cicoli:2011it}:
\begin{equation}
\V =\frac12\, k_{122} t_1 t_2^2= \kappa\, \sqrt{\tau_1}\,  \tau_2\,.
\end{equation}
Here $k_{122}$ is a triple intersection number, $\kappa=\frac{1}{\sqrt{2k_{122}}}$, and we have used the usual relation between 2- and 4-cycle volumes $\tau_i\equiv \partial \V/ \partial t_i$ to find:
\begin{equation}
t_2= \sqrt{\frac{2\tau_1}{k_{122}}}\,,\qquad t_1=\frac{\tau_2}{\sqrt{\tau_1}}\frac{1}{\sqrt{2 k_{122}}}\,.
\end{equation}
Fibred CY threefolds have been extensively used to study cosmological and phenomenological implications of string compactifications (see \cite{Cicoli:2008gp,Burgess:2010bz, Cicoli:2011yy, Broy:2015zba, Cicoli:2016xae, Cicoli:2016chb, Burgess:2016owb, Cicoli:2017axo, Antusch:2017flz, Cicoli:2018cgu, Cicoli:2019ulk, Cicoli:2020bao, Cicoli:2021itv, Cicoli:2022uqa, AbdusSalam:2022krp,Cicoli:2024bxw, Leedom:2024qgr,Cicoli:2024yqh} for an exemplifying list of models). The only thing we are interested in for the moment is the canonical normalisation of the K\"ahler moduli. As stated above, we want to keep the volume fixed and vary only the orthogonal combination of moduli which turns out to be:
\begin{equation}
\label{eq:comb. u}
u = \frac{1}{\kappa} \frac{\tau_1}{\tau_2}\,.
\end{equation}
The volume $\V$ is canonically normalised as in \eqref{eq:can. norm. V}. On the other hand, the canonical normalisation of $u$ reads:
\begin{equation}
u = e^{\sqrt{3}\, \phi/M_p}\,.
\end{equation}
We can thus express the 4-cycle volumes in terms of the overall volume and the $u$ combination as:
\begin{align}
\label{CanNorm}
\tau_1 & = (u \V)^{2/3} = \V^{2/3}\, e^{\frac{2}{\sqrt{3}} \phi/M_p}\,,\\
\tau_2 &= \left(\frac{\V}{\sqrt{u}}\right)^{2/3}= \V^{2/3}\, e^{-\frac{1}{\sqrt{3}}\phi/M_p}\,.
\end{align}
Consequently, the 2-cycle volumes can be expressed as:
\begin{align}
t_1 & \sim \V^{1/3}\, e^{-\frac{2}{\sqrt{3}}\phi/M_p}\,,\\
t_2 & \sim \V^{1/3}\, e^{\frac{1}{\sqrt{3}}\phi/M_p}\,.
\end{align}
Now consider a D3-brane wrapped around the $\mathbb{P}^1$ base which is a 2-cycle with volume $t_1$. Then the tension of the resulting EFT string is given by:
\begin{equation}
\label{eq:tension fibre}
\mu \sim  M_s^2\, e^{-\frac{2}{\sqrt{3}}\phi/M_p}\,,
\end{equation}
which is of the form \eqref{eq:exp. mu} with:
\begin{equation}
\label{eq:xi fibre}
\xi = \frac{2}{\sqrt{3}}\,,
\end{equation}
that satisfies the bound \eqref{eq:lower bound on xi}, and so can in principle accommodate a string growth for certain backgrounds.

On the other hand, if the cycle wrapped by the D3-brane is the 2-cycle of volume $t_2$, the tension is:
\begin{equation}
\label{eq:tension base}
\mu \sim M_s^2\, e^{-\frac{1}{\sqrt{3}}\phi/M_p}\,,
\end{equation}
which has again the form \eqref{eq:exp. mu} with:
\begin{equation}
\xi = \frac{1}{\sqrt{3}}\,,
\end{equation}
which however does not satisfy the bound \eqref{eq:lower bound on xi}. 

Let us now consider the case of an NS5-brane wrapped around the fibre 4-cycle. In this case the relevant volume appearing in \eqref{eq:tension of EFT string} is $\tau_1$. Thus, the tension becomes:
\begin{equation}
\label{eq:mu fibre 4-cycle}
\mu \sim M_s^2\, \tau_1 \sim  e^{\frac{2}{\sqrt{3}}\phi/M_p}\,,
\end{equation}
which is of the form \eqref{eq:exp. mu} with:
\begin{equation}
\label{eq:xi fibre 4-cycle}
\xi = -\frac{2}{\sqrt{3}}\,,
\end{equation}
that meets the growth condition \eqref{eq:lower bound on xi}. An NS5-brane wrapped instead around the 4-cycle with volume $\tau_2$ would not work since one would get $\xi = 1/\sqrt{3}$ which does not satisfy the bound \eqref{eq:lower bound on xi}. 

Note that, in the case of a D3-brane wrapped around the 2-cycle $t_1$, the growth condition (\ref{eq:explicit gc}) becomes:
\begin{equation}
\label{eq:growth region 2-cycle}
X >\frac{1}{\sqrt{2}}\,,
\end{equation} 
which requires a positive $X$. On the contrary, an EFT string arising from an NS5-brane wrapped around the 4-cycle $\tau_1$ can grow in physical size only if $X$ is negative since (\ref{eq:explicit gc}) takes the form:
\begin{equation}
\label{eq:growth region fibre 4-cycle}
X < -\frac{1}{\sqrt{2}}\,.
\end{equation}

Similarly to the case of the volume mode, the leading order effects which develop a potential for fibration moduli tend to be perturbative corrections that scale as:
\begin{equation}
\label{eq:typical pot. tau_f}
V \simeq \frac{\alpha}{\tau_1^q}\,,
\end{equation}
where $\alpha$ depends on $\mathcal{V}$ and other microscopic parameters. Hence, after canonical normalisation using (\ref{CanNorm}), the potential is in general expected to take the exponential form (\ref{eq:exp potential}) with $V_0=\alpha\, \mathcal{V}^{-2q/3}$ and $\lambda = q \, \xi$. We can therefore adapt the results of Tab. \ref{tab:gc fixed points} to express the conditions on $\lambda$ and $\omega$ for which the fixed points of the autonomous system \eqref{eq:X'}-\eqref{eq:Y'} lie within the growth regions \eqref{eq:growth region 2-cycle}-\eqref{eq:growth region fibre 4-cycle}. 

\paragraph*{\textbf{Kination:}} The kination fixed points in the growth region are $\K_+$ in the case of an EFT string arising from a D3-brane on $t_1$, and $\K_-$ for EFT strings which are NS5-branes on $\tau_1$.

As stated in Sec. \ref{sec:volume}, even if kination is not the final attractor of the evolution of the system, for initial conditions close enough to the kination fixed point, there can be a rather long period of cosmic string growth during a transient before the system exits the growth region. The evolution of the system is identical to that of Fig. \ref{fig:evolution}, while the number of e-foldings of growth with respect to the distance from the kination point is very similar to the one shown in Fig. \ref{fig:Ng-Xi}. 

\paragraph*{\textbf{Modulus domination:}}  When the string tension is controlled by the 2-cycle volume $t_1$ as in (\ref{eq:tension fibre}), the fixed point $\mathcal M$ is inside the growth region \eqref{eq:growth region 2-cycle} if:
\begin{equation}
\sqrt{3}< \lambda < \sqrt{6}\,,
\end{equation}
which corresponds to:
\begin{equation}
\frac{3}{2}< q< \frac{3}{\sqrt{2}}\,,
\end{equation}
in the potential \eqref{eq:typical pot. tau_f}. Conversely, when the string tension is controlled by the 4-cycle volume $\tau_1$ as in (\ref{eq:mu fibre 4-cycle}), the fixed point $\mathcal M$ lies within the growth region if:
\begin{equation}
\label{eq:lambda C fibre 4-cycle}
-\sqrt{6}<\lambda<-\sqrt{3}\,,
\end{equation}
corresponding to a range of $q$:
\begin{equation}
\label{eq:q C fibre 4-cycle}
-\frac{3}{\sqrt{2}}<q<- \frac{3}{2}\,.
\end{equation}
In this case $q$ has to be negative, implying that the potential \eqref{eq:typical pot. tau_f} should be an increasing function of $\tau_1$, as expected from the fact that the system must evolve towards smaller values of $\tau_1$ in order for the string to grow.

\begin{figure}[t!]
\centering
\includegraphics[width=0.5\textwidth]{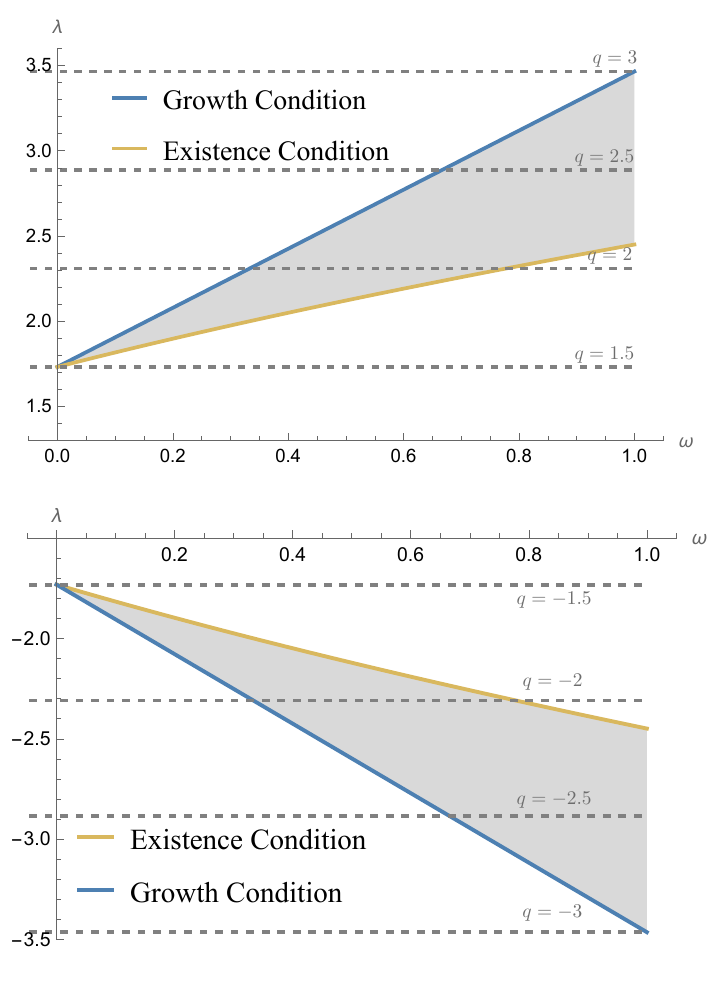}
\caption{Range of $\lambda$ as a function of $\omega$ for which the scaling fixed point lies within the growth region. \emph{Top:} Case of a D3-brane wrapped around the 2-cycle $t_1$; \emph{Bottom:} Case of an NS5-brane wrapped on the 4-cycle $\tau_1$. The dashed lines mark reference values for $q$.}
\label{fig:lambda-omega-fibe}
\end{figure}

\paragraph*{\textbf{Scaling fixed point:}} For the case of a D3-brane wrapped around the base $\mathbb{P}^1$, the scaling fixed point $\mathcal S$ is within the growth region \eqref{eq:growth region 2-cycle} if:
\begin{equation}
\sqrt{3(\omega+1)}< \lambda < \sqrt{3} (\omega+1)\,,
\end{equation}
corresponding to the range of powers:
\begin{equation}
\frac{1}{2} \sqrt{\omega +1}< q < \frac{1}{2}(\omega+1)\,.
\end{equation}
Imposing that the parameter range \eqref{eq:lambda S fibre 4-cycle} exists, the equation of state parameter $\omega$ is constrained to take strictly positive values:
\begin{equation}
\label{eq:omega S fibre 4-cycle}
\omega >0\,.
\end{equation}
On the other hand, in the case of an NS5-brane wrapped around the K3 or $T^4$ fibre, the scaling fixed point $\mathcal S$ lies within the growth region \eqref{eq:growth region fibre 4-cycle} if:
\begin{equation}
\label{eq:lambda S fibre 4-cycle}
- \sqrt{3} (\omega+1)<\lambda < - \sqrt{3(\omega+1)}\,,
\end{equation}
corresponding to a value of $q$:
\begin{equation}
\label{eq:q S fibre 4-cycle}
-\frac{1}{2}(\omega+1)<q<-\frac{1}{2}\sqrt{\omega+1}\,,
\end{equation}
which, again, can exist only if $\omega >0$. The ranges of $\lambda$ with respect to $\omega$ for both cases are plotted in Fig. \ref{fig:lambda-omega-fibe}.

\subsection{Blow-up cycles}
\label{sec:blowup}

Let us now turn to the scenario in which EFT strings arise from wrapping branes around blow-up cycles. These rigid cycles are a key ingredient of LVS moduli stabilisation \cite{Balasubramanian:2005zx, Conlon:2005ki, Cicoli:2008va} and have been extensively exploited in the construction of 4D phenomenological string models (see \cite{Conlon:2005jm, Cicoli:2011qg,Cicoli:2012vw, Cicoli:2013cha, Cicoli:2013mpa, Cicoli:2017shd, Cicoli:2021dhg, Bansal:2024uzr} for illustrative examples). In what follows we shall focus on CY volumes of the form: 
\begin{equation}
\label{eq:Swiss Cheese}
\V = f_{3/2}(\tau_i)- \gamma_s \tau_s^{3/2}\,,
\end{equation}
where $f_{3/2}(\tau_i)$ is a homogeneous function of degree $3/2$ in the 4-cycle moduli $\tau_i$, while $\tau_s$ is a small rigid 4-cycle, i.e. $\tau_s^{3/2} \ll  f_{3/2}(\tau_i)\simeq \V$, and $\gamma_s$ is an $\O(1)$ constant which depends on the triple self-intersection number of the blow-up mode. We shall be interested in the case where the modulus $\tau_s$ is rolling while all the other moduli are fixed. Moreover, we will consider again two different kinds of EFT strings: D3-branes wrapped around the 2-cycle $t_s \sim \sqrt{\tau_s}$ dual to $\tau_s$, and NS5-branes wrapped around the blow-up 4-cycle $\tau_s$. In the first case the resulting 4D EFT string has a tension:
\begin{equation}
\label{eq:mu BU 2cycle}
\mu \sim M_s^2 \, t_s\,.
\end{equation}
The canonical normalisation of $\tau_s$ is given by:
\begin{equation}
\label{eq:can.norm blow-up}
\tau_s = \left(\frac{3 \V}{4 \gamma_s}\right)^{2/3} \left(\frac{\sigma}{M_p}\right)^{4/3}\,,
\end{equation}
and so the tension $\mu$ can be written in terms of $\sigma$ as:
\begin{equation}
\mu \sim M_s^2 \left(\frac{3 \V}{4 \gamma_s}\right)^{1/3} \left(\frac{\sigma}{M_p}\right)^{2/3}\,.
\label{Tension}
\end{equation}
On the other hand, when the whole blow-up 4-cycle is wrapped by an NS5-brane, the tension looks like:
\begin{equation}
\label{eq:tension blowup 4cycle}
\mu \sim M_s^2 \, \tau_s = M_s^2\left(\frac{3 \V}{4 \gamma _s}\right)^{2/3} \, \left(\frac{\sigma}{M_p}\right)^{4/3}\,.
\end{equation}
Note that in both cases the tension is not exponential in the dynamical field $\sigma$. This means that we have to adapt and generalise the machinery developed in Sec. \ref{sec:exp. tension} to study the growth condition. First of all, let us redefine:
\begin{equation}
\label{eq:xi non exp.}
\frac{\xi(\phi)}{M_p} \equiv - \,\frac{\mu,_{ \phi}}{\mu}\,.
\end{equation}
With this definition, $\xi$ is in general field-dependent except for the case of an exponential canonical normalisation. In the present case we obtain $\xi(\sigma)<0$ with:
\begin{equation}
\label{eq:xi blow-up}
|\xi(\sigma)| = r\left(\frac{M_p}{\sigma}\right)\simeq r \sqrt{\frac{\V}{\tau_s^{3/2}}}\gg 1\,,
\end{equation}
where $r=2/3$ for D3-branes on $t_s$, and $r= 4/3$ for NS5-branes on $\tau_s$. Plugging $\xi(\sigma)=-|\xi(\sigma)|$ in the growth condition \eqref{eq:gc xi-phi'}, we find that cosmic strings can grow only if $\sigma$ decreases, as expected from the $\sigma$-dependence of the tensions (\ref{Tension}) and (\ref{eq:tension blowup 4cycle}). Moreover, given that $|\xi(\sigma)|\gg 1$, the lower bound (\ref{eq:lower bound on xi}) is automatically satisfied, implying that the system can in principle lead to a string growth.

This is guaranteed for the kination regime which exists independently of the scalar potential and can always be ensured by an appropriate choice of initial conditions. On the other hand, the existence of a period of string growth for backgrounds different from kination depends on the scalar potential which in this case is however not exponential. This is due to the fact that the canonical normalisation (\ref{eq:can.norm blow-up}) is power-law. In fact, the total potential for the canonically normalised blow-up mode $\sigma$ is in general of the form (setting again $M_p=1$) \cite{Bansal:2024uzr}:
\begin{equation}
\label{eq:total potential blowup}
V\simeq V_0 \left(1- A\, e^{-k \sigma^{4/3}}+ B \, e^{-2k\sigma^{4/3}}- \frac{C}{\sigma^{2/3}}\right),
\end{equation}
where $k\sim \V^{2/3}$ and $A$, $B$ and $C$ depend on $\mathcal{V}$ and on the details of the compactification. The terms proportional to $A$ and $B$ are generated by E3-instantons or gaugino condensation on D7-branes wrapped around the blow-up mode, the term proportional to $C$ is a string loop correction, and the constant is the uplifting needed to achieve a Minkowski vacuum. The potential has a rich structure, featuring a minimum and a plateau region for large field values. If $C$ is sufficiently small, as expected from a detailed analysis of loop corrections \cite{Bansal:2024uzr}, the perturbative term is subleading close to the minimum located at $\sigma=\langle\sigma\rangle> \sigma_c\simeq C^{3/2}$. At the right of the minimum, where $\sigma > \langle\sigma\rangle$, non-perturbative effects become quickly negligible and the potential behaves as $V\simeq V_0\left(1- C\,\sigma^{-2/3}\right)$ before it can be approximated very well by just the constant term for $\sigma\gg \langle\sigma\rangle$. Clearly, the structure of the potential (\ref{eq:total potential blowup}) is more involved than the simple exponential form (\ref{eq:exp potential}). We can therefore generalise the definition of $\lambda$ as follows:
\begin{equation}
\label{eq:lambda general}
\frac{\lambda(\phi)}{M_p} \equiv -\frac{V,_\phi}{V}\,,
\end{equation}
implying that for the potential (\ref{eq:total potential blowup}), $\lambda$ turns out to be field dependent. Thus, in our case the system of equations \eqref{eq:X'}-\eqref{eq:Y'} is no longer closed. This makes the analytic study of the dynamical system impossible, and one has to rely on numerical simulations. One can however still use the results of Tab. \ref{tab:gc fixed points} considering the corresponding fixed points just as approximate and instantaneous. 

For $\sigma\gg \langle\sigma\rangle$, the potential (\ref{eq:total potential blowup}) becomes $V\simeq V_0$, and so $\lambda \to 0$. In this case the results in Tab. \ref{tab:gc fixed points} indicate that the only instantaneous fixed point which can yield a string growth is $\mathcal{K}_-$. When $\sigma> \langle\sigma\rangle$, $|\lambda(\sigma)|$ increases to values of $\mathcal{O}(1)$, and so in principle both $\mathcal{M}$ and $\mathcal{S}$ can exist and give a growth. However numerical studies reveal that these instantaneous fixed points disappear very quickly, leading to at most $\mathcal{O}(0.01)$ e-foldings of growth. In the region close to the minimum, where $\sigma\simeq \langle\sigma\rangle$, non-perturbative effects become important, and so the potential is very steep. Effectively then $|\lambda(\sigma)|\gg 1$ and the only approximated fixed point which can cause a string growth is again $\K_-$. However no relevant string growth happens around the minimum since $\sigma'<0$ only for half oscillation around the minimum and, above all, kination would rapidly turn into matter domination. Again numerical simulations show that close to the minimum one can have only at most $\mathcal{O}(0.01)$ e-foldings of growth.

Thus, we focus on the plateau region where $\sigma\gg \langle\sigma\rangle$ and start close to $\mathcal{K}_-$ by choosing initial conditions such that $X=-1+\Delta$ with $\Delta\ll 1$. This might be ensured if higher order effects induce a steepening of $V$ at field values beyond the plateau region as in \cite{Cicoli:2013oba, Pedro:2013pba}. During kination, the velocity of the field is approximately constant around $\sigma'\simeq-\sqrt{6}$, and so, as long as the system is in the kination regime, $\sigma$ evolves approximately as $\sigma (N) \simeq\sigma_0-(1-\Delta)\sqrt{6} N$. As can be seen from the canonical normalisation (\ref{eq:can.norm blow-up}), $0\leq \sigma \lesssim \mathcal{O}(0.1)$ from the K\"ahler cone conditions. Imposing therefore $\sigma(N)\geq 0$ and $\sigma_0\lesssim \mathcal{O}(0.1)$, we find a strict upper bound on the duration of an epoch of string growth during kination:
\begin{equation}
N_g \lesssim \frac{\sigma_0}{\sqrt{6}} < \mathcal{O}(0.1)\,,
\end{equation}
showing that no significant growth can be achieved for EFT strings arising from branes wrapped around blow-up modes, regardless of the choice of initial conditions.

\section{Conclusions}
\label{sec:conclusions}

Cosmic superstrings with a time dependent tension feature a new, built in mechanism for the formation of string networks. Such networks can lead to interesting observational effects like a stochastic background of gravitational waves than can carry imprints of high scale physics.

In this work we have systematically explored the cosmological dynamics that can give rise to growing strings extending the original work of \cite{Conlon:2024uob}. We have focussed on the case where the string tension is set by a dynamical modulus evolving in the presence of a barotropic fluid under the assumption that the strings do not backreact on the background evolution.

Using the tools of autonomous dynamical systems \cite{Copeland:1997et}, we have shown that a string growth can happen whenever the system lingers for some time in a particular, model dependent region of its phase space, given by \eqref{eq:gc x}, that can include kinating backgrounds \cite{Conlon:2024uob} but also scaling and modulus dominated fixed points.

We have analysed the distinct cases of fundamental and EFT strings. For fundamental strings, the tension is set by the string mass which depends on both the volume modulus and the dilaton. We have shown that only a dynamical volume modulus may give rise to a string growth, and that this growth may take place both as a fast-roll transient or at a scaling fixed point. We demonstrated that the existence and duration of the growth phase depends on the initial conditions, the scalar potential and the barotropic fluid. 

EFT strings appear at low energies when a $p$-brane wraps a $(p-1)$-cycle. In these cases the string tension depends on the volume of the wrapped cycle. Focusing on type IIB compactifications, we have analysed the cases with $p=3$ and $p=5$, and identified the bulk cycles of fibred CY threefolds as the most promising candidates. More specifically, we have shown that the growth condition can be satisfied for a D3-brane wrapping the base 2-cycle and for an NS5-brane wrapping the fibre 4-cycle. We have constrained the slope of the potential and the equation of state parameter of the barotropic fluid. Finally we have analysed D3 and NS5-branes wrapping blow-up cycles. Though this setup evades an exact analytical treatment, given that neither the potential nor the string tension are of exponential form, we were able to show that growth is in principle possible for kinating backgrounds, provided the initial field velocity and displacement can be chosen to satisfy the field dependent growth condition. However, the duration of this growth phase turns out to be too short to yield an appreciable effect. The reason is the fact that blow-up modes typically feature sub-Planckian field ranges that cannot sustain an enduring kination epoch.

The analysis of the fixed points of the dynamical system singles out particular values of the equation of state parameter $\omega$ and the moduli dependence of the scalar potential ($p$ for the volume mode, and $q$ for fibration moduli) which can cause a string growth. It is therefore important to check if these parameters can assume realistic values. In particular, for $\V$, the physically interesting cases for $p$ involve: $p=4/3$ for $\overline{{\rm D3}}$-branes at the tip of a warped throat \cite{Kachru:2003sx}, $p=2$ for $\overline{{\rm D3}}$-branes in a CY region where warping is negligible \cite{Kachru:2003aw} or non-zero F-terms of the dilaton or the complex structure moduli \cite{Gallego:2017dvd}, $p=8/3$ for logarithmic redefinitions of $\tau_b$ \cite{Conlon:2010ji} or T-branes \cite{Cicoli:2015ylx}, and $p=3$ for $\alpha'$ corrections \cite{Becker:2002nn}. On the other hand, for $\tau_1$, string loops and $F^4$ $\alpha'$ corrections can give $q\in \{-2,\,-1,\,-1/2,\,1/2,\,1\}$ (see for example \cite{Cicoli:2008gp,Cicoli:2016chb}). In a given explicit model, it is then important to check which power of $p$ for $\V$ and $q$ for $\tau_1$ give the dominant contribution to $V$ in the region where the relevant modulus is rolling.

A fundamental point left open by this work is the cosmological mechanism leading to the production of an initial population of strings. A precise understanding of this production mechanism is crucial for assessing the number density of the initial population of strings. This would yield a lower bound on the number of e-foldings of growth the strings should undergo before meeting each other and eventually percolating into a cosmic network. Combining this information with the present analysis may further constrain the underlying dynamics and lead to sharper phenomenological predictions. 

Thanks to our results, the growth of the physical size of cosmic strings looks more general. It is worth stressing that this has implications not just for the formation of a network from an initial population of isolated small loops or for the evolution of an already formed network, but also for a gas of highly excited strings which is the other option at the end of brane-antibrane inflation \cite{Frey:2005jk,Frey:2023khe}. In fact, a growth of the physical size of the strings in the gas might potentially alter the duration of the Hagerdon phase or the spectrum of emitted gravity waves \cite{Frey:2024jqy}. We leave this study for future work.

A case our work does not touch upon are field-theoretic cosmic strings formed from the spontaneous breaking of a gauge group that contains a $U(1)$-factor \cite{Vilenkin:2000jqa}. The tension of these strings depends on the vacuum expectation value of the Higgs field responsible for the symmetry breaking. As noted in \cite{Revello:2024gwa}, in type II string compactifications where gauge groups live on branes wrapped around cycles, the value of the Higgs field would depend on the modulus controlling the volume of the wrapped cycle and on the open string moduli living on the brane. A complete assessment of the dynamics and growth of these strings is another interesting direction for future work.

\subsection*{Acknowledgments}

We would like to thank Ivano Basile, Joseph P. Conlon, Vittorio Larotonda and Gonzalo Villa for useful discussions. The work of LB, MC and FGP contributes to the COST Action COSMIC WISPers CA21106, supported by COST (European Cooperation in Science and Technology).

\bibliography{references}

\end{document}